\documentclass[a4paper,aps, prl,amsmath,amssymb, superscriptaddress, showpacs, twocolumn, floatfix]{revtex4}
\usepackage{graphicx, braket, mathrsfs, nicefrac, epsfig, multirow, ulem, verbatim, subfigure}

\begin{document}
\title{Phenomenology and Dynamics of Majorana Josephson Junction}
\author{D. I. Pikulin}

\affiliation{Instituut-Lorentz, Universiteit Leiden, P.O. Box 9506, 2300 RA Leiden, The Netherlands}

\author{Yuli V. Nazarov}

\affiliation{Kavli Institute of Nanoscience, Delft University of Technology, Lorentzweg 1,
2628 CJ Delft, The Netherlands}

\begin{abstract}
We derive a generic phenomenological model of a Majorana Josephson junction that accounts for avoided crossing of Andreev states, and investigate its dynamics at constant bias voltage to reveal an unexpected pattern of any-$\pi$ Josephson effect in the limit of slow decoherence.
\end{abstract}
\maketitle

Recently, the proposals of solid-state realizations of Majorana fermions came into focus of attention. While the first proposal \cite{Moore} concerned non-Abelian excitations in $5/2$ FQHE in semiconductor heterostructures, most proposals \cite{Kitaev,Ivanov} exploited exotic superconductors where Majorana fermions correspond to zero-energy states of an effective BdG Hamiltonian. The Majorana states are instrumental for realization of topological quantum computation \cite{Kitaev1}.
More recent contributions \cite{BeenakkerReview} utilize the proximity effect from a conventional superconductor, either in nanowires in a strong magnetic field and with strong spin-orbit interaction \cite{Sau,Lutchyn,Oreg1}, or in topological insulators \cite{Fu0, Akhmerov}. This brings the Majoranas close to experimental realization, and underlines the importance of reliable experimental signatures of their presence. Among the signatures are half-integer conductance quantization \cite{Wimmer} and $4\pi$ Josephson effect in superconductor-superconductor (SS) junctions\cite{Fu2, Lutchyn, Oreg}.

No $4\pi$ periodicity is to be seen in the stationary ground state of the junction
It can only be observed \cite{Houzet, von Oppen} in dynamics induced, for instance, by a d.c. voltage bias. Unambiguous signature of this anomalous periodicity is the current noise peak at half of the Josephson frequency $\omega_j =2eV/\hbar$ \cite{Houzet}. We have suggested that the avoided crossing of Andreev states is intrinsic for finite systems and restores the $2\pi$ periodicity of the junction ground state \cite{Us}. This has been recently confirmed by detailed calculations of the Andreev spectrum of the nanowire-based SS junctions \cite{Aguado}.
\begin{figure}
\centerline{\includegraphics[width=0.9\linewidth]{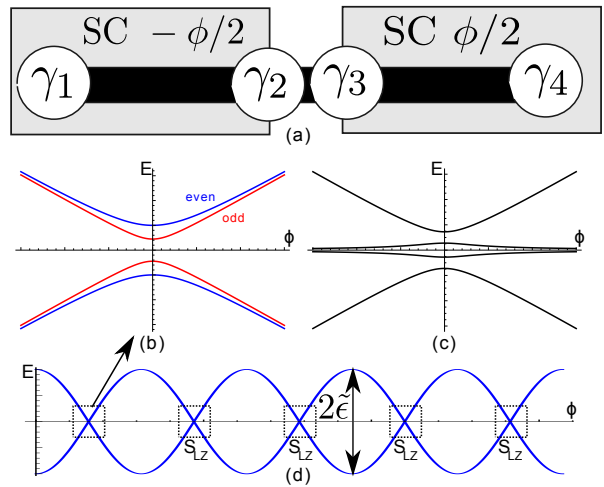}}
\caption{a. A Majorana Josephson junction is formed by mounting a nanowire (black) on two superconducting leads (grey) resulting in four Majoranas $\gamma_{1-4}$.  b. The energies of the junction states versus phase near an avoided crossing point. c. Corresponding Andreev levels. d. The energies at bigger phase scale.
}
\label{fig:setup}
\end{figure}

In this Letter, we put forward a generic phenomenological model of a Majorana Josephson junction and  demonstrate that the dynamics in the junction are substantially richer than thought. In particular, the sharp peaks in noise spectrum of a voltage-biased junction are not generally confined to any definite fraction of $\omega_j$: one can talk of $any-\pi$ Josephson effect in this context. Experimental observation of these singularities would give a robust proof of the existence of Majoranas and open up the possibilities for quantum manipulation of these states. Our treatment of dynamics encompasses the Landau-Zener tunneling at the avoided crossings, decoherence, relaxation, and quasiparticle poisoning. 

We exemplify with a nanowire setup (Fig. \ref{fig:setup}) although the same phenomenology extends to topological insulators. A nanowire mounted on a single superconducting lead develops a topologically non-trivial state in a parameter range of magnetic fields and gate voltages\cite{Lutchyn}. Two Majorana states emerge at the wire ends. Majorana Josephson junction is formed by mounting the wire on two leads biased with superconducting phase difference $\phi$. Two extra Majorana states $\gamma_{2,3}$ emerge at the junction, in addition to the end states $\gamma_{1,4}$. The overlap between $\gamma_2$ and $\gamma_3$ is strong but does depend on phase and vanishes at a certain $\phi_0$. If one disregards the end states \cite{Fu2, Lutchyn, Oreg}, the resulting energies are $4\pi$ periodic in $\phi$ and the resulting states are of indefinite parity. We exemplify this dependence with $E(\phi) = \pm \tilde{\epsilon} \sin(\frac{\phi-\phi_0}{2})$. To fix the parity, it is paramount to bring the end states to the picture. We developed \cite{Us} a scattering matrix theory where the $2\pi$ periodicity is proven from the topological properties of the scattering matrix. In a nutshell, the crossing of Andreev levels is avoided. We need a practical Hamiltonian to describe the details of the situation  in the vicinity of $\phi_0$. That can be rigorously derived from the scattering approach, yet we opt here for a simple heuristic deviation in terms of overlaps of Majorana states.

  These overlaps are exponentially small for long wires, $\propto \exp(-L/2\xi)$, $L$ being the wire length, the localization length $\xi$ being of the order of the spin-orbit length $L_{so}$. For InAs wires \cite{Kouwenhoven}, $L_{so} = 0.2 \mu m$, and $L$ would not exceed $2 \mu m$ since inevitable disorder forbids topological state for longer wires. This sets the biggest exponential suppression to $\simeq 10^{-2}$.  Owing to the exponential suppression, the direct overlap $t_{14}$ between the end states is much smaller than the overlaps between the end and the junction states, and can be disregarded. This brings us to the following Hamiltonian:
\begin{align}
\hat{H} = \tilde\epsilon(\phi - \phi_0) \hat{\gamma}_2\hat{\gamma}_3+ \notag \\
(t_{12}\hat{\gamma}_2 + t_{13}\hat{\gamma}_3)\hat{\gamma}_1 +
(t_{42}\hat{\gamma}_2 + t_{43}\hat{\gamma}_3)\hat{\gamma}_4.
\label{eq:Hamiltonian}
\end{align}
that is valid in the vicinity of the crossing point and provides a generic phenomenological model of a Majorana Josephson junction. Here, the overlaps $t$ are real, and $\hat{\gamma}_{1-4}$ are self-conjugated anticommuting Majorana operators \cite{Majorana}.

It is instructive to give the eigenenergies of the full many-body states of the Hamiltonian, rather than the associated Andreev levels. The Hamiltonian conserves the parity of the particle number and therefore gives rise to eigenstates with either odd or even number of particles. There are two eigenvalues of opposite sign for each parity,
\begin{equation}
E_{\pm,{\rm o,e}} = \pm \sqrt{G^2_{{\rm o,e}} + \frac{\tilde{\epsilon}^2}{4} (\phi-\phi_0)^2},
\end{equation}
where 
\begin{align}
G_{o,e} =\tfrac{1}{2}\sqrt{(t_{12}\pm t_{43})^2 +(t_{13} \mp t_{42})^2}
\end{align}
and $\pm$ sign is chosen such that $G_e > G_o$.
Their phase dependence (Fig. \ref{fig:setup}b)gives a familiar glimpse of avoided level-crossing hyperbolas,
$G_{\rm o,e}$ being the minimum energy splittings of odd/even states, respectively.
The two positive energies  of the associated Andreev levels are given by $E_{1,2} = |E_{{\rm e}}| \pm |E_{{\rm o}}|$ (Fig.\ref{fig:setup}c).  This characteristic form is conformed by the numerical calculations based on microscopic models \cite{Us, Aguado} proving the validity of the Hamiltonian (\ref{eq:Hamiltonian}). The absence of the direct overlap $t_{14}$ leads to a special property: the phase-dependent term describing the overlap of $\hat{\gamma}_2$, $\hat{\gamma}_3$ anticommutes with the rest of the terms. This guarantees the energies to be even in phase and to merge far from $\phi =\phi_0$, these properties would be absent for a most general four-Majorana Hamiltonian. 

\begin{figure}
\centerline{\includegraphics[width=0.7\linewidth]{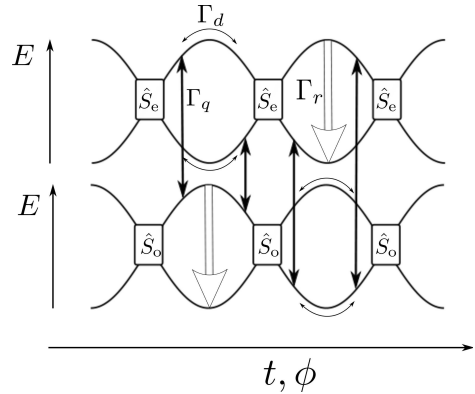}}
\caption{Processes affecting the junction dynamics. In between the parity-dependent Landau-Zener scatterings (described by $2\times2$ matrices $\hat S_{\text{o},\text{e}}$) the junction is subject to dephasing (with a rate $\Gamma_d$), relaxation ($\Gamma_r$) and quasiparticle poisoning ($\Gamma_q$).}
\label{fig:dynamics}
\end{figure}

Let us notice that the junction in either odd or even state is nothing but a qubit that is similar to other superconducting qubits that commonly exploit avoided level crossing\cite{Pashkin, Shumeiko, Gorelik}. One can employ quantum manipulation of the resulting Majorana states by changing the superconducting phase in time. For instance, following \cite{Pashkin}, one can prepare the qubit in the ground state reasonably far from the crossing point $\phi=\phi_1$, and give a pulse that brings the junction to $\phi=\phi_0$. This will cause Rabi oscillations with frequency $2G/\hbar$ that can be detected by measuring the probabilities to find the qubit in the ground or excited state after the pulse as functions of pulse duration. 

Here we restrict ourselves to the case of immediate experimental relevance where the junction is biased by a d.c. voltage $V$ so that the phase $\phi$ is swept linearly with time, $\dot{\phi} = 2eV/\hbar$.  In a usual Josephson junction where the energy levels are $2\pi$ periodic, such bias results in coherent oscillations of the supercurrent $I(\phi)= 2e/\hbar \partial E(\phi)/\partial \phi$ with Josephson frequency $\omega_j = 2eV/\hbar$ \cite{Josephson}. The idea behind the $4\pi$ Josephson effect \cite{Kane} is an apparent $4\pi$ periodicity of energy levels in the limit of vanishing $G$, this suggests the oscillations at  a half of Josephson frequency, $I(t) = \pm I_{\text{m}} \cos(\omega_jt/2)$, $I_{\text{m}} \equiv e\tilde{\epsilon}/\hbar$. Albeit these oscillations  cannot be coherent owing to random switching between the two branches $\pm$ of the energy spectrum. The signature of $4\pi$ periodicity is rather a sharp peak in the spectral density of the current {\it noise} \cite{Houzet}, with the width of order of  switching rate, and integrated intensity being given by $I_{\text{m}}^2/2$, the average square of the current. For this simplified picture to hold, one should require sufficiently small voltages, $V \ll \tilde{\epsilon}$. Failure to satisfy this condition results in proliferation to higher energy levels and finally to continuous spectrum, this increasing the peak width to the values of the order of $\Delta$ and thus rendering noise peaks undetectable. \cite{Houzet,Aguado}

In this Letter, we address the noise in Majorana Josephson junctions at smaller voltages. Evidently,
the avoided level crossing results in usual Josephson effect in the limit $V\to 0$. The complex and interesting crossover between $2\pi$ and $4\pi$ regimes involves Landau-Zener (LZ) tunneling
 upon crossing a point $\phi=\phi_0 + 2\pi n$ in the vicinity of the point. 
 The parity obviously does not change, and for each parity we have a classic setup of LZ tunneling \cite{LandauZener} between two levels. The values of the qubit wave function 
before and after LZ scattering are related by a $2\times2$  unitary matrix:
\begin{equation}
\hat S_{o, e} = \left(\begin{array}{cc}
\sqrt{1-P_{o, e}} & - e^{i\chi} \sqrt{P_{o, e}}\\
e^{-i\chi} \sqrt{P_{o, e}} &  \sqrt{1-P_{o, e}}
\end{array}\right), \label{eq:LZ}
\end{equation}
where the  probability of LZ tunneling is given by 
\begin{eqnarray}
P_{o, e} = \exp\left(-\frac{4\pi}{eV}\frac{G_{o, e}^2}{\tilde \epsilon}\right).
\end{eqnarray}
This suggests an importance of a voltage scale $eV_0 \equiv \simeq 4\pi G^2/\tilde{\epsilon} \ll \tilde{\epsilon}$ at which the probabilities are of the order of $1$ and the crossover between $2\pi$ and $4\pi$ regimes is expected. We stress that the probabilities are generally different for odd and even states that permits the identification of these states that are hardly distinguishable otherwise. 

The quantum dynamics are affected by the processes of relaxation, dephasing and quasiparticle poisoning (Fig.\ref{fig:dynamics}) that occur throughout the time-line with no peculiarities near the crossing points. We assume low temperature $k_B T \ll \tilde{\epsilon}$, so that the relaxation is always from higher to lower energy state with the rate $\Gamma_r(\phi)$. The decoherence suppresses the non-diagonal elements of the density matrix (with the rate $\Gamma_{d}(\phi)$) not affecting the diagonal ones. We assume the fluctuation of the phase $\phi$ to be the main source of the decoherence, in this case $\Gamma_d(\phi) \propto I^2(\phi)$. The quasiparticle transfer processes account for a parity change. They may be due to stray quasiparticles in the bulk superconductor that come to the junction with the energies of the order of the superconducting energy gap $\Delta > \tilde{\epsilon}$ and lose this energy either adding or annihilating a quasiparticle in Andreev levels under consideration. Due to significant initial quasiparticle energy, the probabilities to find the junction in either upper or lower state after a quasiparticle transfer, are the same. The quasiparticle rate $\Gamma_{q}$ does not depend on the phase $\phi$.

\begin{figure}
\centerline{\includegraphics[width=0.9\linewidth]{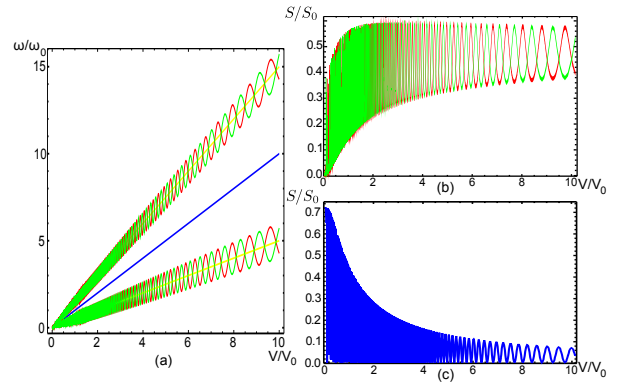}}
\caption{The slow decoherence limit. We chose $\tilde{epsilon}/eV_0 =30$ for all plots. (a). Frequency positions of the noise peaks versus $V$ ($\omega_0=(2e/\hbar)V_0$). The main ones are at $n\omega_j$ while the positions of additional peaks oscillate converging at $(n+1/2)\omega_j$. Only $n=0,1$ are shown. (b) Integrated noise intensity (in units of $S_0 \equiv I^2_{\text{m}}/2$) of the first two additional peaks. (c) The same for the $n=1$ main peak.}

\label{fig:results}
\end{figure}

This results in the straightforward but lengthy equation for the density matrices $\hat{\rho}_{o,d}$ to be found in \cite{supplementary}. We solve this equation with continuity conditions 
$\hat{\rho}_{o,d}(t_{ac}+0) = \hat{S}_{o,d} \hat{\rho}_{o,d}(t_{ac} - 0) \hat{S}^{-1}_{o,d}$, $t_{ac}$ corresponding to time moments of the crossings, and compute the correlator of current operators 
\begin{equation}
S(\omega) =\int^{\infty}_{-\infty} d t_1 \int^{\frac{2\pi}{\omega_j}}_0 d t_2 e^{i\omega (t_1-t_2)}\langle\langle I(t_1) I(t_2)\rangle\rangle. \label{eq:S} 
\end{equation}
that gives the spectral density of the current noise.
\begin{figure}
\centerline{\includegraphics[width=0.9\linewidth]{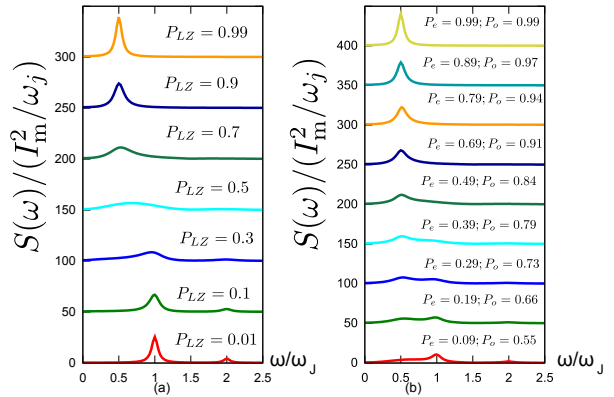}}
\caption{
The spectral intensity $S(\omega)$ of the current noise for a set of $V$ corresponding to LZ probabilities shown, in the limit of fast decoherence.
a. Indistinguishable parities $P_e=P_o=P_{LZ}$; b. $P_e\neq P_o$ ($G_{\text{e}} = 4 G_{\text{o}}$). 
Distinct peaks at multiples of $\omega_j$ at $P \ll 1$
transmute into a single peak at $\omega_j/2$ at $P\approx 1$.
($\Gamma_{r}=\Gamma_{q} = 0.02 \omega_j$ for all plots).}
\label{fig:peaks}
\end{figure}
We concentrate on two limiting cases of fast (Fig. \ref{fig:peaks}) and slow (Fig. \ref{fig:results})decoherence. In both cases, we assume slow relaxation and poisoning, $\omega_j \gg \Gamma_r,\Gamma_{q}$. 

"Fast" implies the quantum coherence is lost during a period of the Josephson oscillations, $\Gamma_d \gg \omega_j$, and the equation for density matrix reduces to a master equation. 
Fig. \ref{fig:peaks}a shows the spectral density for equal LZ probability for even and odd states, $G_{e}\approx G_{o}$.
The voltage growth from the lowermost to upper curve resulting in increased $P_{LZ}$. At low voltage ($P_{LZ} \ll 0$, the noise peaks at $\omega_j$ as well as at its multiples, the latter manifesting non-sinusoidal $I(\phi)$. This proves a usual periodicity.
At higher voltage where $P_{LZ} \approx 1$ we see a single peak at $\omega_j/2$ manifesting $4\pi$ periodicity.
In both limiting cases, the peak widths $\simeq \Gamma_{r,qp}$.
Important feature is the absence of any distinguishable peaks at intermediate $P_{LZ}$. The reason is the LZ tunneling causing incoherent switching at almost any crossing point. The peaks acquire width $\simeq \omega_j \gg \Gamma_{r,qp}$ and correspondingly reduce their height to the background level. 

In Fig. \ref{fig:peaks}b the LZ probabilities are very different at the crossover corresponding to $G_{e}/G_{o} =4$. Now one can distinguish the peaks at both $\omega_j/2$ and $\omega_j$ in the crossover region, though they are reduced in height in comparison with the limiting cases. The explanation is the parity separation in time domain. Since $\Gamma_{q}\ll \omega_j$, the parity persist over many periods between the random switches. While the junction is in even parity state, $P_{LZ} \approx 1$, and during this time interval the noise at $\omega_j/2$ is generated. While the junction is in odd parity state, almost no LZ tunneling takes place, and the noise is generated at Josephson frequency. The experimental observation of two peaks would thus prove the parity effect. One can also think of a more challenging observation where the noise can be resolved fast, that is, at a time-scale $< \Gamma^{-1}_{q}$. Such noise measurement will monitor the parity of the junction in real time.

The results in the opposite limit of slow decoherence $\Gamma_d\ll \omega_j$ are decisively more complex and intriguing (Fig. \ref{fig:results}). In this limit, the dynamics are truly quantum over many periods. An analytical analysis gives the positions of the noise peaks as well as the integrated noise intensities around each peak\cite{supplementary}. Most striking feature is an oscillatory dependence of the peak intensities and positions on voltage. This is a manifestation of quantum interference between the subsequent LZ tunneling events not suppressed by decoherence. Similar interference patterns have been predicted and observed for superconducting qubits in\cite{MachZender,Gorelik}. We have found that a voltage-biased Majorana Josephson junction presents the simplest and most striking framework for this interference effect.

The quantum phase $\theta$ accumulated between the subsequent crossing points is estimated as 
\begin{align}
\theta = \mathop{\int}_{\text{period}} dt \frac{\Delta E(\phi(t))}{\hbar} = \frac{8\tilde{\epsilon}}{\hbar \omega_j}
\label{eq:theta}
\end{align}
The phase is big on the scale $eV/\epsilon$, its increment by $2\pi$ gives an estimate of the oscillation period in voltage $\Delta V = (\pi/8) V(eV/\tilde{\epsilon})\ll V$. 

Importantly, the frequency positions of the additional noise peaks (Fig. \ref{fig:results}a), those the main Josephson peaks at multiples of $\omega_j$, are not at any integer fractions of $\omega_j$. In the context, we can dub this any-$\pi$ Josephson effect. It stems from a quasi-energy splitting in a periodically driven qubit. At $V \gg V_0$, additional peaks converge at $(2n+1)\omega_j/2$ oscillation around this frequency. The spread of these oscillations $\Delta \omega$ does not vanish with increasing $V$: rather, it increases following $\Delta \omega \simeq (2e/\hbar)\sqrt{VV_0}$. This proves that any-$\pi$ Josephson effect can be observed at voltages $V\gg V_0$ far beyond the crossover region. The width of the peaks is determined by $\Gamma_d$. From this, we estimate the minimum decoherence rate permitting the resolution of the peaks: $\Gamma_d \simeq (e/\hbar)\sqrt{V V_0}$. For the sake of simple drawing,  we assumed  indistinguishable
 parities such that $P_{\text{o}} = P_{\text{e}}$.
If $P_{\text{o}} \ne P_{\text{e}}$, the additional peaks split once again corresponding to the two parities.\cite{supplementary}

At $V\ll V_0$, the noise intensity is mainly concentrated at a main peak at $\omega_j$. In the opposite limit, the intensity concentrates at the peaks converging to $\omega_j/2$ retaining oscillating features even at high voltage. (Fig.\ref{fig:results}b,c)

To summarize, we have derived a generic phenomenological Hamiltonian to describe a Majorana Josephson junction with avoided Andreev level crossing, and investigated its quantum dynamics at constant voltage bias with emphasis on noise signatures of the anomalous periodicity. While in the fast decoherence regime the signatures follow an expected pattern, the interference of the subsequent LZ tunneling events results in a complex any-$\pi$ Josephson effect pattern in slow decoherence regime. The experimental observation of the effects predicted will provide unambiguous signature of Majorana states in Josephson junction and open up the perspectives of quantum manipulation and parity measurements in such junctions.

This research was supported by the Dutch Science Foundation NWO/FOM. The authors are indebted to C. W. J. Beenakker, L. P. Kouwenhoven and especially to R. Aguado for useful discussions.

\begin{widetext}
\section{Appendix}

\subsection{Periodic continuation of the phenomenological Hamiltonian}

Far from the avoided crossing point ($|\phi-\phi_0| \gg G/\tilde{\epsilon} \ll 2\pi$), the energies merge together, so that the lowest Andreev level is close to zero,  $E_2 \simeq G^2/\tilde{\epsilon} \ll G \ll \tilde{\epsilon}$. At further increase of $|\phi-\phi_0|$,  the energies deviate from their linear asymptotes near the crossing points to become  $2\pi$ periodic(Fig. \ref{fig:setup}d). We approximate their dependence with 
\begin{equation}
E_o=E_e=\pm\tilde{\epsilon}\left|\sin\left(\frac{\phi-\phi_0}{2}\right)\right|. 
\end{equation}  
It may seem that the Hamiltonian may be extended to the full range of the phases simply by replacing $\phi-\phi_0$ with $2\sin((\phi-\phi_0)/2))$. However, this would result in a $4\pi$-periodic  Hamiltonian, this being at odds with  the natural $2\pi$ periodicity. 

To resolve this apparent discrepancy, we note that the choice of Majorana operators in Eq. \ref{eq:Hamiltonian} that describes the lowest energy states,  is not unique: one can substitute instead of $\gamma_{1-4}$ any linear combinations of Majorana operators of a bigger set that obey the commutation relations. The members of this bigger set would correspond to higher energy states not considered. 
The choice of four Majorana operators $\hat{\gamma}_{1-4}$ made does not depend on phase in the vicinity of the avoided crossing point and does depend at a bigger scale.  The choice  in fact is a $4\pi$ periodic one. 
To give an example of a $2\pi$-periodic Hamiltonian, let us substitute $\gamma_3$ in the form of  two Majorana operators that do not depend on the superconducting phase,
\begin{equation}
\hat{\gamma}_3 =\cos((\phi-\phi_0)/2)\hat{\gamma}'_3 +  \sin((\phi-\phi_0)/2) \hat{\gamma}''_3
\end{equation}

With this, we can rewrite a seemingly $4\pi$ periodic term in the form
\begin{eqnarray}
\tilde{\epsilon} \sin((\phi-\phi_0)/2) \hat{\gamma}_2  \hat{\gamma_3} 
 \to \frac{\tilde{\epsilon}}{2} \left((1-\cos(\phi-\phi_0))\hat{\gamma}'_3 + \sin(\phi-\phi_0) \hat{\gamma}''_3\right)
\end{eqnarray}
which makes the $2\pi$ periodicity explicit. In further considerations, we will work in a basis of the energy eigenstates that is explicitly $2\pi$ periodic. This makes irrelevant the details of Majorana representation outlined here. 

\subsection{Equation for density matrix}
Let us give here the evolution equations on density matrix that are straightforward but too bulky to fit into the main text.

Beyond the vicinities of the crossing points $\phi = \phi_0 + 2\pi n$, $n$ being an integer, we can disregard the terms
proportional to $G_{\text{o},\text{e}}$. We work in local eigenenergy basis. To denote the elements of the density matrix,
we use the subscripts $e$ and $o$ for even and odd parity sectors, respectively, and $u$ and $l$ for upper or lower states in each parity sector.
We collect all the incoherent processes: relaxation, dephasing and quasiparticle poisoning. We introduce energy splitting $ E(\phi(t)) = 2\tilde{\epsilon}\left|\sin\left(\frac{\phi-\phi_0}{2}\right)\right|$.  With this, the equations read: 
\begin{eqnarray}
\frac{d\rho_{eu, eu}}{dt} =& 
-  \Gamma_r\rho_{eu, eu} -  \Gamma_q \rho_{eu, eu}+ \tfrac{1}{2} \Gamma_q\left( \rho_{ou, ou}+\rho_{ol, ol}\right) \\
\frac{d\rho_{el, el}}{dt} = & 
 \Gamma_r\rho_{eu, eu} -  \Gamma_q \rho_{eu, eu}+ \tfrac{1}{2} \Gamma_q\left(\rho_{ou, ou} + \rho_{ol, ol}\right) \\
\frac{d\rho_{eu, el}}{dt} = &
\left(-i E(\phi(t)) -\Gamma_d- \Gamma_q -\tfrac{1}{2} \Gamma_{r}\right) \rho_{eu, el},\\
\frac{d\rho_{el, eu}}{dt} = &\left(i E(\phi(t))  - \Gamma_d - \Gamma_q -\tfrac{1}{2} \Gamma_{r}\right)\rho_{el, eu},\\
\frac{d\rho_{ou, ou}}{dt} = &
-   \Gamma_r\rho_{ou, ou} -  \Gamma_q\rho_{ou, ou} + \tfrac{1}{2} \Gamma_q\left( \rho_{eu, eu} +  \rho_{el, el} \right) \\
\frac{d\rho_{ol, ol}}{dt} =  &
\rho_{ou, ou} \Gamma_r - \rho_{ou, ou} \Gamma_q +\tfrac{1}{2} \Gamma_q\left( \rho_{eu, eu} + \rho_{el, el} \right) \\
\frac{d\rho_{ou, ol}}{dt} = &
\left(-i E(\phi(t)) -\Gamma_d- \Gamma_q -\tfrac{1}{2} \Gamma_{r}\right) \rho_{ou, ol}\\
\frac{d\rho_{ol, ou}}{dt} =  &
\left(i E(\phi(t))  - \Gamma_d - \Gamma_q -\tfrac{1}{2} \Gamma_{r}\right)\rho_{ol, ou}.
\end{eqnarray}

To specify the $\phi$ dependence of the decoherence rate, we assume that the decoherence takes place mainly due to the non-ideal bias conditions that give rise to the fluctuations of $\phi$.
Those cause the fluctuations of the energy splitting $\delta E =(\partial E/\partial \phi) \delta \phi$ and are converted to the fluctuations of quantum phase thereby. In this case, $\Gamma_d \propto (\partial E/\partial \phi)^2$ and can be thus chosen to be of the form $\Gamma_d(\phi) = \Gamma_d \cos^2((\phi-\phi_0)/2)$.

In the vicinities of the crossing points we may neglect the incoherent terms. For each parity, the Hamiltonian (\ref{eq:Hamiltonian})can be written as a $2\times2$ matrix
$$
\hat{H}= \left[ \begin{array}{cc} G_{\text{o,e}} & \tilde{\epsilon}(\phi-\phi_0)\cr
-\tilde{\epsilon}(\phi-\phi_0) &  -G_{\text{o,e}} \end{array}\right]
$$
Integrating this over time and transforming to the eigenenergy basis reproduces the boundary condition on the density matrix given in the main text, for density matrices $\rho_{b}$($\rho_a$) before (after) passing the crossing point,
\begin{equation}
\rho_a = \hat{S}\rho_b\hat{S}^{-1};
\end{equation}
where the scattering matrix of LZ tunneling is given by Eq. \ref{eq:LZ} and $\chi = \frac{\pi}{4} + \mathrm{Arg}\left[\Gamma(1- i \frac{G^2}{eV\tilde \epsilon} )\right] + \frac{G^2}{eV\tilde \epsilon} (\ln \frac{G^2}{eV\tilde \epsilon} - 1) $ ($\Gamma$ here is the gamma function) 

\subsection{Fast decoherence limit and master equation}
Let us first consider the case of the fast decoherence when the quantum coherence quenches at the time scale smaller than the period of Josephson oscillations, $\Gamma_d \gg \omega_j$. This permits us to take into account only the diagonal elements of density matrix, the probabilities, describe their evolution with a master equation. We denote the probabilities with $p_{\alpha, \beta}$, where $\alpha = e, o$ at even or odd sectors and $\beta = u, l$ for upper or lower state in the sector. With all the processes  in Fig. \ref{fig:dynamics} taken into account, the master equation for time intervals between the crossings reads:
\begin{equation}
\frac{dp_{eu}}{dt} = - (\Gamma_r + \Gamma_q) p_{eu} + \Gamma_q\frac{p_{ou} + p_{ol}}{2},
\end{equation}
\begin{equation}
\frac{dp_{el}}{dt} = \Gamma_r p_{eu} - \Gamma_q p_{el}  + \Gamma_q\frac{p_{ou} + p_{ol}}{2},
\end{equation}
\begin{equation}
\frac{dp_{ou}}{dt} = - (\Gamma_r + \Gamma_q) p_{ou} + \Gamma_q\frac{p_{eu} + p_{el}}{2},
\end{equation}
\begin{equation}
\frac{dp_{ol}}{dt} = \Gamma_r p_{ou} - \Gamma_q p_{ol}  + \Gamma_q\frac{p_{eu} + p_{el}}{2}.
\end{equation}
This has to be supplemented by LZ boundary conditions at the crossing points for the probabilities before and after the passing. For the odd sector, we have
\begin{align}
p^{\text{a}}_{ol} = p^{\text{b}}_{ol} + P_{\text{o}} \left(p^{\text{b}}_{ou}-p^{\text{b}}_{ol}\right);\;
p^{\text{a}}_{ou} = p^{\text{b}}_{ou} + P_{\text{o}} \left(p^{\text{b}}_{ol}-p^{\text{b}}_{ou}\right).
\end{align}
Similar equation holds for the probabilities in the even sector. 

Let us label the four possible states $ou,ol,eu,el$ with a single index $j$
Solution of these equation in the long time limit approaches $p^{(0)}(t)_j$, that is periodic in time with the Josephson period. To compute the correlator of the currents, we also need the propagator of the evolution equation $U_{ij}(t_2,t_1)$. It is defined at $t_2 > t_1$ as the solution of the equation at the time moment $t_2$, $p_i(t_2)$, with initial condition 
$p_{k}(t_1) = \delta_{kj}$.

The current is a function of a state given by
\begin{equation}
I_i(t) = I_{\text{m}} \text{sgn}(\phi-\phi_0) \cos((\phi-\phi_0)/2)\left(\delta_{i,ou} +\delta_{i,eu} - \delta_{i,ol} -\delta_{i,el}\right)
\end{equation}
with $I_{\text{m}} = (2e/\hbar) (\tilde{\epsilon}/2)$,
$\phi=\phi_0 +\omega_j t$. The correlator
is expressed as
\begin{equation}
\langle\langle I(t_1) I(t_2)\rangle\rangle = 
\sum_{ij}  I_i(t_1) I_{j}(t_2) (U_{ij}(t_1,t_2)-p^{0}_i(t_1)) p^{(0)}_j(t_2), 
\end{equation}
at $t_1>t_2$, and is obtained by permutation of the time arguments otherwise.

We solve the equation, the propagator and find the correlator numerically. The results for the current noise spectral density are presented in Fig. \ref{fig:peaks}. 

\subsection{Details of slow decoherence limit}

Interesting analytical results can be obtained in the opposite limit of slow decoherence such that $\Gamma_{d,r,q} \ll \omega_j$ if in addition we assume that $\text{max}\left(P_{LZ},\Gamma_q/\omega_j\right) \gg \Gamma_r/\omega_j$ (the latter condition even in the absence of $\Gamma_q$ is satisfied at $V > V_0 /\ln(\Gamma_r/\omega_j$ and thus certainly holds in the crossover regime). Under these assumptions, the relaxation can not set a preferential state. All possible states of the junction are present with equal probability, and long time limit density matrices do not depend on time and approach $\hat{\rho}_{o}=\hat{\rho}_{e}=\tfrac{1}{4} \hat{1}$. 
In this case, we can neglect $\Gamma_{d,r,q}$ implement the pure quantum dynamics to compute the current-current correlator at time separation $|t_1-t_2| \ll \Gamma^{-1}_{d,r,q}$. While not enough to resolve fine features of the noise spectral density such as the line-shapes of the noise peaks, this suffices to evaluate the integrated noise intensities in the vicinity of each peak. 

The positions of the peaks are not bound to the multiples or integer fractions of the Josephson frequency. To understand this in general, let us note that the independent solutions $|\Psi_j\rangle$ of the Schr\"{o}dinger equation that is in our case periodic with the Josephson period $T_j \equiv 2\pi/\omega_j$, are Bloch-like functions of time satisfying 
\begin{equation}
|\Psi_k(t+ T_j)\rangle = \exp(i\lambda_k) |\Psi_k(t)\rangle
\end{equation}
and having Fourier components at discrete frequencies 
$\omega_j (n + \lambda_k/2\pi)$. The correlators thus can have Fourier components at all discrete frequencies satisfying
$\omega_j(n +(\lambda_k -\lambda_l)/2\pi)$.

To analyze the quantum dynamics in the case under consideration it is proficient to apply a unitary transform that cancels the evolution of the wave function during the "free motion" between the crossing points. The phase difference $\chi$ accumulated in the course of the free motion (Eq.\ref{eq:theta} ) is then ascribed to the LZ scattering matrix in certain parity sector so it becomes
\begin{equation}
\hat{S}= \left(\begin{array}{cc}
\sqrt{1-P} e^{i\theta/2} & - e^{i\chi} \sqrt{P}\\
e^{-i\chi} \sqrt{P} &  \sqrt{1-P} e^{-i\theta/2}
\end{array}\right), 
\end{equation} 
Since this matrix describes the evolution of the wave function over the period, its eigenvalues give $exp(-i\lambda)$ and
\begin{equation}
\cos(\lambda) = \sqrt{1-P}\cos(\theta/2).
\end{equation}
Let us define $\gamma \equiv \arccos(\sqrt{1-P}\cos(\theta/2))/\pi$, $0<\gamma<1$,
and recall that we have two parity sectors and correspondingly two parameters $\gamma_{\text{o},\text{e}}$.
With this, the noise spectral density is digested in the form that makes the peak positions explicit:
\begin{align}
S(\omega) = \sum_{n>0} S_n \tfrac{1}{2}\left(\delta (\omega - n \omega_j) + \delta (\omega + n \omega_j)\right) + \notag \\
\sum_{n\ge0} S^{+,\text{o}}_n \tfrac{1}{2}\left(\delta (\omega -( \gamma_{\text{o}} + n) \omega_j) + \delta (\omega +( \gamma_{\text{o}} + n) \omega_j)\right) + \notag \\
\sum_{n>0,\pm} S^{-,\text{o}}_n \tfrac{1}{2}\left(\delta (\omega -( -\gamma_{\text{o}} + n) \omega_j) + \delta (\omega +( -\gamma_{\text{o}} + n) \omega_j)\right) + \notag \\
\sum_{n\ge0} S^{+,\text{e}}_n \tfrac{1}{2}\left(\delta (\omega -( \gamma_{\text{e}} + n) \omega_j) + \delta (\omega +( \gamma_{\text{e}} + n) \omega_j)\right) + \notag \\
\sum_{n>0,\pm} S^{-,\text{e}}_n \tfrac{1}{2}\left(\delta (\omega -( -\gamma_{\text{e}} + n) \omega_j) + \delta (\omega +( -\gamma_{\text{e}} + n) \omega_j)\right). \notag 
\end{align}

The integrated intensities $S_n$ of the main peaks are contributed by both parity sectors, while  the additional peaks $S_n^{\pm,\text{o/e}}$ are in general resolved in parity with respect to the positions and height.
It is worth noting that the definition of noise density in use is "quantum", so that the noise at positive and negative frequencies does not have to be the same indicating the difference between emission and absorption of quanta with energy $\hbar \omega$. However, this is not the case in the present framework: the calculation explicitly gives the spectral density that is even in frequency.

To compute the intensities, we need to evaluate the correlator in Eq. \ref{eq:S}. We note that in the representation used
\begin{equation}
\hat{I}(t) = I_{\text{m}} \cos(\omega_j t/2) \hat{\sigma}_z;
\end{equation}
(crossing points corresponding to $t = n T_j$),
and the evolution matrix $\hat{U}(t_2,t_1)$ that gives the wave function at $t_2$ from the initial condition at $t_1$ is simply given by
\begin{equation}
\hat{U}(t_2,t_1) = \left(\hat{S}\right)^{N},
\end{equation}
$N$ being the number of the crossing points at the interval $(t_2,t_1)$.

For the intensities of the main peaks, this gives
\begin{equation}
S_{n} = I^{2}_m \frac{16 n^2}{\pi^2\left(4n^2-1\right)^2} \sum_{\text{o},\text{e}} \frac{1}{2}\left(1 - \frac{P_{\text{o},\text{e}}}{\sin^2(\theta/2)+P_{\text{o},\text{e}}\cos^2(\theta/2)} \right)
\end{equation}

As to the additional peaks, their intensities are given by
\begin{equation}
S(\Omega)=I^{2}_m \left[\frac{4}{\pi}\frac{\cos(\pi\Omega/\omega_j) \Omega\omega_j}{\omega^2_j - 4 \Omega^2}\right]^2\frac{P_{\text{o},\text{e}}}{\sin^2(\theta/2)+P_{\text{o},\text{e}}\cos^2(\theta/2)}
\end{equation}
where $\Omega$ is the (parity-dependent) frequency position of the additional peak, $P$ is the LZ probability at the corresponding parity.

Making use of the above relations, we plot in Fig. \ref{fig:results} the positions and intensities of the three lowest peaks assuming $P_{\text{o}}=P_{\text{e}}$, that hinders parity resolution.

\subsection{Extra figures}
To illustrate the slow decoherence limit in more detail, we present here two extra figures. Fig. \ref{fig:supp1} illustrates the noise intensities of the two peaks converging to $\Omega =(3/2)\omega_j$ and of the second main peak at $\Omega=2\omega_j$. In both cases, the relative intensity is substantial in the crossover region $V\simeq V_0$ and slowly falls off upon increasing $V$. Fig. \ref{fig:supp} represents the generic case of parity separation. The LZ probabilities here are taken to be distinctly different in the crossover region  corresponding to $G_{e} = 4G_{o}$. We see that a so-to-say $4\pi$ periodic peak at $\omega_j/2$ in fact consists of the four distinct peaks of different intensity slowly converging to $1/4$ of the full intensity in the limit $V\gg V_0$.

\begin{figure}
\centerline{\includegraphics[width=0.9\linewidth]{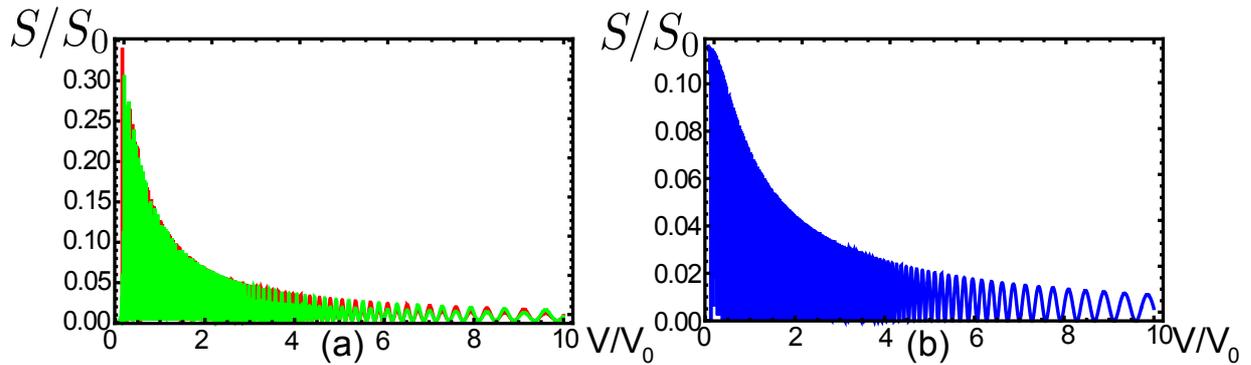}}
\caption{ Intensities of the high-frequency noise peaks. The parameters are the same as in Fig. \ref{fig:results}. Left: two additional peaks with frequencies converging to $(3/2)\omega_j$ at $V \gg V_0$.Right: main peak at $2 \omega_j$. }
\label{fig:supp1}
\end{figure}

\begin{figure}
\centerline{\includegraphics[width=0.9\linewidth]{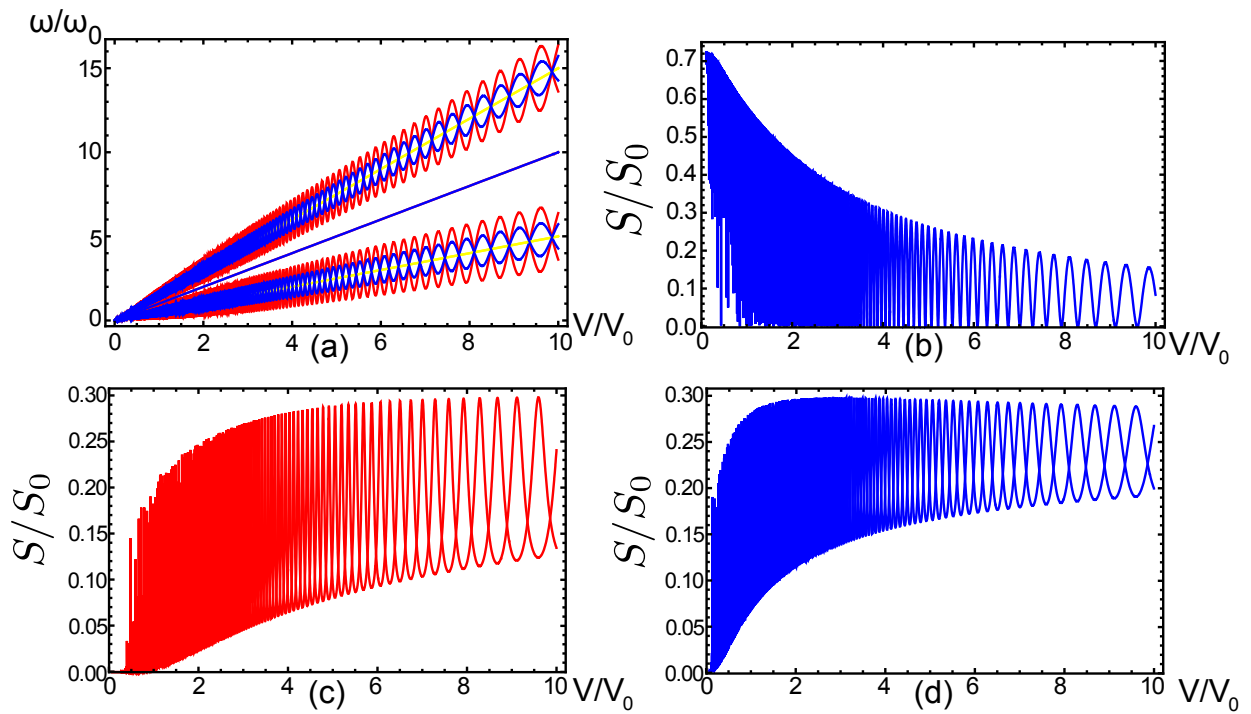}}
\caption{ The generic case of parity separation. Here, $V_0 = 4\pi G^2_{o}/ \tilde{\epsilon}$, $\omega_0 = 2e V_0/\hbar$, $\tilde{\epsilon}/\omega_0 =30$, $S_0 \equiv I_{\text{m}}^2/2$. 
a. Peak positions. Four distinct noise peaks converge to half-integer multiples of $\omega_j$ upon increasing voltage.
b. Spectral intensity of the main peak at $\omega_j$. It is contributed by both parities. c. The same for the first two even parity peaks. d. The same for the  two  odd parity peaks. }

\label{fig:supp}
\end{figure}
\end{widetext}

\end{document}